\documentclass[epj]{svjour}

\usepackage[utf8]{inputenc}
\usepackage[english]{babel}
\usepackage{graphicx}
\usepackage{physics}
\usepackage{amsmath}
\usepackage{amssymb}
\usepackage{bbm}
\usepackage{tikz}
\usepackage{arydshln}
\usepackage{multirow}

\usepackage{xcolor}
\usepackage{slashed}
\usepackage{hyperref}
\usepackage{placeins}
\newcommand{\K}[1]{\ensuremath{\left(#1\right)}}
\newcommand{\Ke}[1]{\ensuremath{\left[#1\right]}}

\allowdisplaybreaks
\begin{document}
\title{Towards Hypernuclei from Nuclear Lattice Effective Field Theory}
\author{Fabian Hildenbrand\inst{1} \and Serdar Elhatisari\inst{2,3} \and{Zhengxue Ren}\inst{1} \and{Ulf-G. Mei\ss ner\inst{3,1} }
}                     
\institute{Institute~for~Advanced~Simulation, Forschungszentrum~J\"{u}lich, 
D-52425~J\"{u}lich,~Germany\and Faculty of Natural Sciences and Engineering, Gaziantep Islam Science and Technology University, Gaziantep 27010, Turkey
\and   Helmholtz-Institut für Strahlen- und Kernphysik and Bethe Center for Theoretical Physics, Universität Bonn, 53115 Bonn, Germany}
\date{Received: date / Revised version: date}
%
\abstract{
Understanding the strong interactions within baryonic systems beyond 
the up and down quark sector is pivotal for a comprehensive description 
of nuclear forces. This study explores the interactions involving hyperons, 
particularly the $\Lambda$ particle, within the framework of nuclear lattice effective 
field theory (NLEFT). By incorporating  $\Lambda$ hyperons into the NLEFT framework, 
we extend our investigation into the $S = -1$ sector, allowing us to probe the 
third dimension of the nuclear chart. We calculate the  $\Lambda$ separation energies 
($B_{\Lambda}$) of hypernuclei up to the medium-mass region, providing valuable insights 
into hyperon-nucleon ($YN$) and hyperon-nucleon-nucleon ($YNN$) interactions. 
Our calculations employ high-fidelity chiral interactions at N${}^3$LO for nucleons 
and extend it to  $\Lambda$ hyperons with leading-order S-wave $YN$ interactions as well as
$YNN$ forces constrained only by the $A=4,5$ systems. Our results contribute to a deeper understanding of the SU(3) symmetry breaking and establish a foundation for future improvements
in hypernuclear calculations.
} 

\maketitle

\section{Introduction}

Understanding the strong interactions in the light quark sector is crucial 
for a comprehensive description of baryonic systems. In the up and down quark 
(nucleonic - $N$) sector, the strong interactions can be described by highly 
successful phenomenological potential models based on meson field theory and dispersion 
relations, such as Paris $NN$~\cite{Lacombe:1980dr}, Stony Brook $NN$~\cite{Jackson:1975be}, 
Nijmegen I-II $NN$~\cite{Stoks:1994wp}, AV18 $NN$~\cite{Wiringa:1994wb}, CD~Bonn~$NN$~\cite{Machleidt:2000ge}, 
Urbana-Argon $NN$ and $3N$~\cite{Pieper:2001mp}, and Urbana-Illinois $3N$~\cite{Pudliner:1997ck} potentials or 
by the framework of chiral effective field theory ($\chi$EFT)~\cite{Epelbaum:2008ga,Machleidt:2011zz} which is considered 
the modern theory of the nuclear 
forces between nucleons. To probe the strong interactions beyond the up and down quark sector, hyperons, 
such as the $\Lambda$ particle, offer a unique opportunity by extending the traditional nuclear chart 
into the third dimension, combining a hyperon with a nucleus to form hypernuclei.
Since the Pauli exclusion principle does not apply between nucleons and hyperons, in this new type of atomic nuclei the $\Lambda$ separation energy, $B_{\Lambda}$, can exceed the binding energy per nucleon in conventional
nuclear systems, although the latter is larger in light hypernuclei.

The study of hypernuclei provides valuable insights into the baryon-baryon interactions, 
and an accurate description of the properties of hypernuclei requires a systematic formulation 
of interactions between hyperons and nucleons, as well as constraining their low-energy constants (LECs). 
The great success of both phenomenological potential models and chiral EFT for nucleons is based on 
rich and precise $NN$-scattering data and nuclear binding energies. However, due to the scarcity of 
hyperon-nucleon and hyperon-hyperon scattering data, the spectra of hypernuclei are pivotal in 
constraining these interactions, deepening our understanding of SU(3) flavor 
symmetry breaking and charge symmetry breaking in strong interactions.

There has been intense interest and significant progress in the study of 
hypernuclei from both theoretical and experimental programs. For a 
comprehensive review of past and recent efforts, see, for example, 
Ref.~\cite{Gal:2016boi}. Early theoretical work on medium-mass hypernuclei has 
been explored using the shell model \cite{Gal:1972gd,Auerbach:1981mq} 
and phenomenological models \cite{Dalitz:1972vzj,Millener:1988hp,Kumar:2023ryb}. 
Calculations for larger hypernuclear systems typically employ the $G$-matrix method 
and Skyrme Hartree-Fock approaches \cite{Yamamoto:1988qz,Haidenbauer:2019thx,Fernandez:1989zw,Lanskoy:1997xq,Tretyakova:1999wv,Cugnon:2000gm,Vidana:2001rm,Zhou:2007zze}, as well as relativistic mean-field models \cite{Lu:2014wta,Meng:2016book,Rong:2021bim,Ding:2022gbu}

One of the leading theoretical efforts to investigate hypernuclei is the No-Core-Shell-Model (NCSM), 
which describes light hypernuclei with great precision using interactions derived 
within $\chi$EFT~\cite{Le:2023bfj,Le:2022ikc,Liebig:2015kwa,Wirth:2017lso,Wirth:2017bpw,Wirth:2016iwn,Wirth:2014apa}.
However, this method is currently not suited for addressing the medium- and heavy-mass regions due to 
computational scaling.
Light hypernuclei have also been explored using cluster models \cite{Hiyama:2009ki,Hiyama:2012sq,Hiyama:2013owa}.
Additionally, the light mass region has been studied within the framework of pionless effective field theory \cite{Schafer:2022une,Hildenbrand:2019sgp,Contessi:2018qnz,Contessi:2019rxv}. 
Furthermore, initial studies of hypernuclear systems using quantum Monte Carlo (MC) calculations have been performed 
in the medium and heavy mass regions \cite{Lonardoni:2017uuu,Lonardoni:2013gta}.

Experimentally, hypernuclei are often produced through reactions like 
($K^-$, $\pi^-$)~\cite{CERN-Heidelberg-Warsaw:1973cbk,Bruckner:1976st,Heidelberg-Saclay-Strasbourg:1981wxg}, 
($\pi^+$, $K^+$)~\cite{Milner:1985kr,Pile:1991cf,Hasegawa:1996fj}, 
and electromagnetic processes ($e,e^{\prime}K^+$)~\cite{HNSS:2002max,JeffersonLabE91-016:2004qei}. 
Early experiments using ion collisions at GeV energies 
demonstrated the feasibility of producing light hypernuclei~\cite{Kerman:1973zz}, with subsequent experiments 
at facilities 
like Dubna~\cite{Bando:1988gb} and GSI~\cite{Ekawa:2022qjt,Saito:2023fnx} further advancing the field. 
Modern experiments at J-PARC~\cite{J-PARCE40:2022nvq,Mbarek:2021vja} and the ALICE experiment at 
LHC~\cite{ALICE:2019eol,ALICE:2022yyh} as well as BESIII~\cite{BESIII:2023clq,Zhang:2024ene}
continue to investigate hyperon-nucleon interactions.

In this work, we study hypernuclei system in the framework of nuclear lattice 
effective field theory (NLEFT). NLEFT is a powerful quantum many-body method that combines 
the advantages of effective field theories with lattice methods~\cite{Lee:2008fa,Lahde:2019npb}. 
It offers the unique feature that the computational time scales 
only with the mass number $A$. The method has recently been used to compute the ground state energies, excited 
state energies and charge radii of light and medium-mass nuclei, and to describe 
the saturation energy and density of symmetric nuclear matter at next-to-next-to-next-to-leading order 
(N$^3$LO) in $\chi$EFT simultaneously 
reproducing accurate two-nucleon phase shifts and mixing 
angles~\cite{Elhatisari:2022qfr}.

In the framework of NLEFT, the inclusion of $\Lambda$ hyperons was first explored 
in Ref.~\cite{Frame:2020mvv} using the impurity lattice Monte Carlo (ILMC) 
method~\cite{Elhatisari:2014lka}. This study employed a simplified Wigner 
SU(4)-symmetric interaction to compute the binding energies of light 
hypernuclei ${}^3_\Lambda$H, ${}^4_\Lambda$H and $^5_\Lambda$He. 
The ILMC method treats single $\Lambda$ hyperons as worldlines in a medium 
of nucleons simulated by the Auxiliary Field Quantum Monte Carlo 
(AFQMC) method. Later, the ILMC method was expanded to 
the study of systems containing multiple hyperons~\cite{Hildenbrand:2022imw}.
Recently, a novel AFQMC approach has been introduced to enable the efficient
calculations of hypernuclear systems with an arbitrary number of hyperons, 
and lattice simulations for pure neutron matter and hyper-neutron matter up to five 
times nuclear matter saturation density have been performed~\cite{Tong:2024egi}.

In this paper, we extend the {\it ab initio} method reported in~\cite{Lee:2008fa,Lahde:2019npb,Elhatisari:2022qfr} 
towards the $S=-1$ sector by including the lightest hyperon, the $\Lambda$, 
and its interactions with nucleons. This allows us to address the third dimension 
of the nuclear chart within this powerful framework. As a starting point we calculate 
$\Lambda$ separation energies of 
hypernuclei up to the medium-mass region in the framework of NLEFT. This work not only advances 
our theoretical understanding but also lays the groundwork for 
future improvements in hypernuclear calculations, ultimately contributing to a deeper comprehension 
of the strong interaction in baryonic systems.
Possible applications in the future include an in depth structure analysis 
of hypernuclei using the pinhole algorithm similar to the work done for the carbon 
nucleus in Ref.~\cite{Shen:2022bak}.

The paper is structured in the following way. We start with a brief overview about the general 
formalism and the underlying interactions in Sec.~\ref{sec:formalism}, at which point we introduce 
the newly included $YN$ and $YNN$ forces, before discussing the fitting procedure of the latter 
ones in Sec.~\ref{sec:fit}. After a brief discussion of finite box size effects in Sec.~\ref{sec:finitebox}, 
we conclude this paper with a detailed presentation and analysis of a selection of  
medium-mass hypernuclei with $A = 3 \ldots 16$ in Sec.~\ref{sec:results}. Finally, we 
discuss potential improvements to the methods used here. Technical details are relegated to the appendices.

\section{Lattice Hamiltonian\label{sec:formalism}}

\begin{figure*}[htb]
    \centering
    \includegraphics[width=1.0\textwidth]{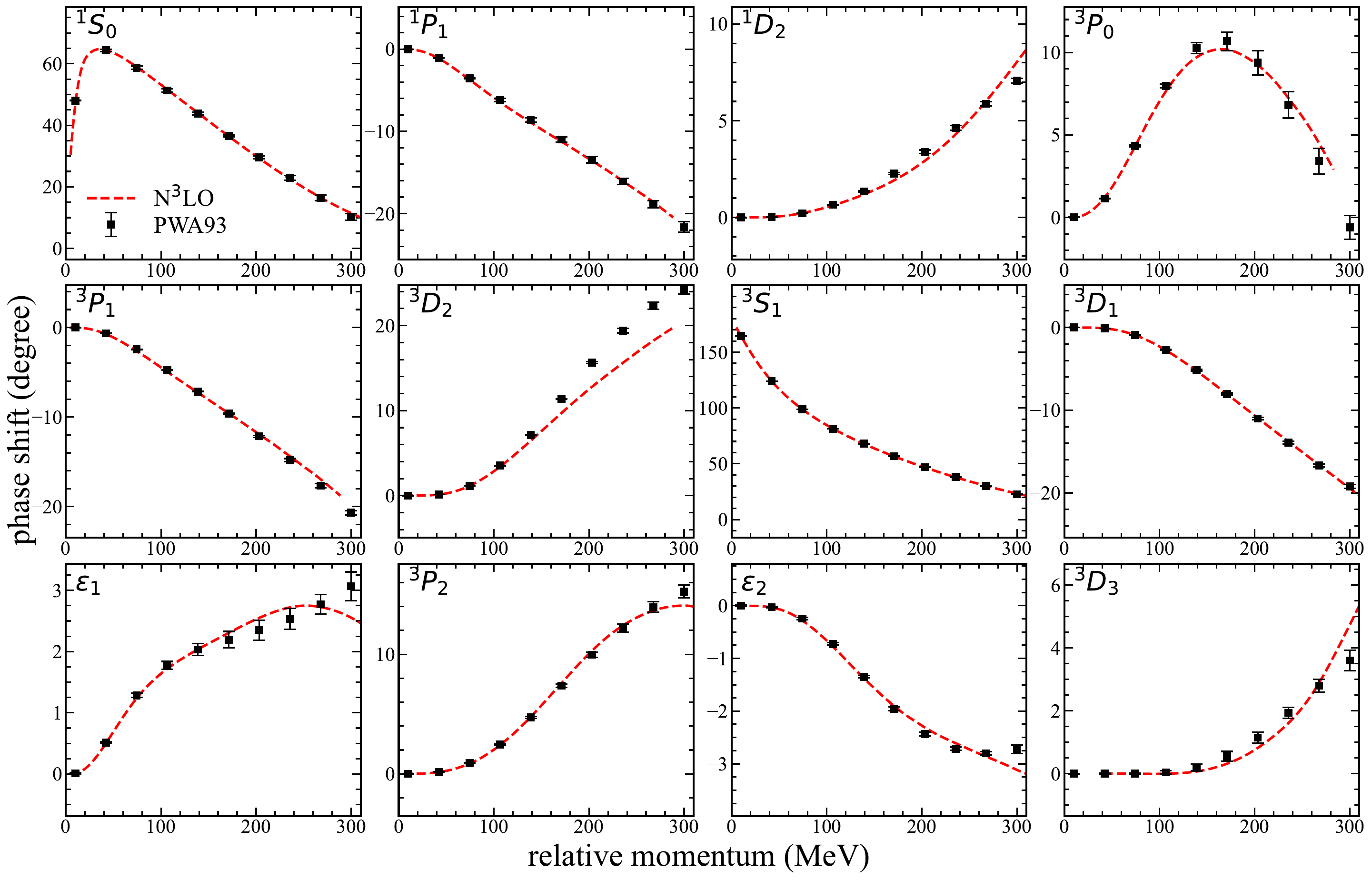}
    \caption{The neutron-proton scattering phase shifts (S,P,D waves)
            and mixing angles ($\epsilon_{1,2}$) up to N$^3$LO in $\chi$EFT versus the relative momentum. The partial wave analysis is taken from Ref.~\cite{Stoks:1993tb}.}
    \label{fig:NN-scattering}
\end{figure*}

We start by constructing and discussing the lattice Hamiltonian needed for 
systems consisting of nucleons and $\Lambda$ hyperons.
In general, two different types of interactions can be defined, the ones involving only nucleons and the interactions containing additional hyperons.
For the former, we use high-fidelity chiral interaction at N$^3$LO and 
the quantum many-body approach, so called wave function matching, developed 
in Ref.~\cite{Elhatisari:2022qfr}. This lattice action has been used to 
successfully describe the spectrum and charge radii of nuclei over 
a large part of the nuclear chart as well as the saturation properties of nuclear 
matter~\cite{Elhatisari:2022qfr}, structure factors for 
hot neutron matter~\cite{Ma:2023ahg}, and the nuclear charge radii of silicon 
isotopes~\cite{Konig:2023rwe} simultaneously 
maintaining accurate two-nucleon phase shifts and mixing 
angles as shown in Fig.~\ref{fig:NN-scattering}. Since 
the lattice operators and lattice Hamiltonian for nucleons are extensively 
discussed in the Supplementary Sections of Ref.~\cite{Elhatisari:2022qfr}, 
we will focus primarily on the lattice operators and interactions involving hyperons.

For the latter ones, in this work, we limit ourselves to leading order 
S-wave $\Lambda N$  contact interactions:
\begin{align}
V_{Y N} =& \frac{1}{4}C_{YN}^S(\mathbbm{1}-\boldsymbol{\sigma_1}\cdot\boldsymbol{\sigma_2})
+\frac{1}{4}C_{YN}^T(3+\boldsymbol{\sigma_1}\cdot\boldsymbol{\sigma_2})~,
\label{eq:V-lambdaN}
\end{align}
and additional contact three-body $\Lambda NN$ forces as derived in Ref.~\cite{Petschauer:2020urh}
\begin{subequations}\label{eq:tbf}
\begin{align}
    V_{YNN} =& C_1(\mathbbm{1}-\boldsymbol{\sigma}_2\cdot\boldsymbol{\sigma}_3)(3+\boldsymbol{\tau}_2\cdot\boldsymbol{\tau}_3)\\
    &+ C_2 \,\boldsymbol{\sigma}_1\cdot(\boldsymbol{\sigma}_2+\boldsymbol{\sigma}_3)(\mathbbm{1}-\boldsymbol{\tau}_2\cdot\boldsymbol{\tau}_3)\\
    &+ C_3 (3+\boldsymbol{\sigma}_2\cdot\boldsymbol{\sigma}_3)(\mathbbm{1}-\boldsymbol{\tau}_2\cdot\boldsymbol{\tau}_3)\,,
\end{align}
\label{eq:V-LambdaNN}
\end{subequations}
where $\boldsymbol{\tau},\boldsymbol{\sigma}$ are Pauli-(iso)spin matrices and $C_i$ are the 
respective LECs. The three-body forces (TBFs) appear at N$^2$LO in the 
chiral power counting \`a la Weinberg. In pionless EFT the three-body forces, however, 
would be leading order. Since here we do not consider the explicit two-pion exchange interactions, 
which are effectively simulated by the smearing discussed below, 
we promote the three-baryon forces in the $S=-1$ sector to LO. 
Note further that the one-pion exchange is suppressed for the $\Lambda N$ 
interaction due to isospin symmetry and hence not part of this effective leading 
order interaction. Finally, due to the fact that high-fidelity chiral interactions between 
nucleons are well tested, 
those set the smearing parameters as well as the lattice spacing for the hypernuclear 
interactions.  Therefore, our lattice Hamiltonian is defined as,
\begin{align}
H = H_{\rm N^3LO} + H^{Y}_{\rm free} + V_{Y N} + V_{YNN}\,,
\label{eq:H-001}
\end{align}
where $H_{\rm N^3LO}$ is the high-fidelity Hamiltonian for nucleons~\cite{Elhatisari:2022qfr}, 
$H^{\Lambda}_{\rm free}$ is the kinetic energy term for $\Lambda$ 
hyperons defined by using fast Fourier transforms to produce the exact dispersion 
relations $E_\Lambda=p^2/(2m_{\Lambda})$ with hyperon mass 
$m_{\Lambda}=1115.68$~MeV, $V_{Y N}$ and $V_{YNN}$ are the hyperon-nucleon 
and hyperon-nucleon-nucleon interactions given in Eqs.~(\ref{eq:V-lambdaN}) 
and (\ref{eq:V-LambdaNN}), respectively.

Before describing $V_{Y N}$ and $V_{YNN}$ interactions, we define 
the total densities for nucleons and hyperons. We begin with the non-smeared 
total nucleon, spin and isospin densities at lattice side $\boldsymbol{n}$ in terms of annihilation (creation) 
operators $a_{i,j}^{}$ ($a_{i,j}^{\dagger}$),
\begin{subequations}
\begin{align}
\rho(\boldsymbol{n})=&\sum_{i,j=0,1}a_{i,j}^\dagger (\boldsymbol{n})a_{i,j}^{} (\boldsymbol{n})\,,
    \\
\rho_S(\boldsymbol{n})=&\sum_{i,j,i^{\prime}=0,1} a_{i,j}^\dagger (\boldsymbol{n})
[\boldsymbol{\sigma}_{S}]_{i,i^{\prime}}a_{i^{\prime},j}^{}(\boldsymbol{n})             
    \\
\rho_I(\boldsymbol{n}) =&\sum_{i,j,j^{\prime}=0,1} a_{i,j}^\dagger (\boldsymbol{n})
[\boldsymbol{\tau}_{I}]_{j,j^{\prime}}a_{i,j^{\prime}}^{} (\boldsymbol{n})
\\
\rho_{SI}(\boldsymbol{n}) =&\sum_{i,j,i^{\prime},j^{\prime}=0,1} a_{i,j}^\dagger (\boldsymbol{n})
[\boldsymbol{\sigma}_{S}]_{i,i^{\prime}}[\boldsymbol{\tau}_{I}]_{j,j^{\prime}}a_{i^{\prime},j^{\prime}}(\boldsymbol{n})\,,
\end{align}
\end{subequations}
and the non-smeared total hyperon and spin densities in terms of 
annihilation (creation) 
operators $b_{i}^{}$ ($b_{i}^{\dagger}$),
\begin{subequations}
\begin{align}
     \xi(\boldsymbol{n}) =& \sum_{i=0,1} b_{i}^\dagger (\boldsymbol{n})b_{i}^{} 
    (\boldsymbol{n})\,,
    \\
     \xi_{S}(\boldsymbol{n}) =& \sum_{i,i^{\prime}=0,1} b_{i}^\dagger (\boldsymbol{n})
                                [\boldsymbol{\sigma}_{S}]_{i,i^{\prime}}
                                 b_{i^{\prime}}^{}(\boldsymbol{n})\,,
\end{align}
\end{subequations}
where $i = 0, 1$ (up, down) denotes the spin index, and $j = 0, 1$ (proton, neutron) 
is the isospin index. Similarly, we define the non-locally smeared operators with $\boldsymbol{\tau}_I$ , $\boldsymbol{\sigma}_S$ the Pauli-matrices in isospin- and spin-space, respectively  ,
\begin{subequations}
\begin{align}
     \hat{\rho}(\boldsymbol{n}) &=\sum_{i,j=0,1}\tilde{a}_{i,j}^\dagger (\boldsymbol{n})
                                     \tilde{a}_{i,j}^{} 
    (\boldsymbol{n})\,,
    \\
     \hat{\rho}_S(\boldsymbol{n}) &=\sum_{i,j,i^{\prime}=0,1} \tilde{a}_{i,j}^\dagger (\boldsymbol{n})
                             [\boldsymbol{\sigma}_{S}]_{i,i^{\prime}}\tilde{a}_{i^{\prime},j}^{} 
    (\boldsymbol{n})\,,
    \\
     \hat{\rho}_I(\boldsymbol{n}) &=\sum_{i,j,j^{\prime}=0,1} \tilde{a}_{i,j}^\dagger (\boldsymbol{n})
    [\boldsymbol{\tau}_{I}]_{j,j^{\prime}}\tilde{a}_{i,j^{\prime}}^{}(\boldsymbol{n})  \,,
    \\
     \hat{\rho}_{SI}(\boldsymbol{n}) &=\sum_{i,j,i^{\prime},j^{\prime}=0,1} \tilde{a}_{i,j}^\dagger (\boldsymbol{n})
    [\boldsymbol{\sigma}_{S}]_{i,i^{\prime}}[\boldsymbol{\tau}_{I}]_{j,j^{\prime}}\tilde{a}_{i^{\prime},j^{\prime}}^{} 
(\boldsymbol{n})\,,
    \\
     \hat{\xi}(\boldsymbol{n}) &= \sum_{i=0,1} \tilde{b}_{i}^\dagger (\boldsymbol{n})
                                     \tilde{b}_{i}^{} (\boldsymbol{n})\,,
    \\
     \hat{\xi}_{S}(\boldsymbol{n}) &= \sum_{i,i^{\prime}=0,1} \tilde{b}_{i}^\dagger (\boldsymbol{n})
                                [\boldsymbol{\sigma}_{S}]_{i,i^{\prime}}
                                \tilde{b}_{i^{\prime}}^{}
    (\boldsymbol{n})\,, 
\end{align}
    
\end{subequations}
\noindent
where $\tilde{a}$ ($\tilde{a}^{\dagger}$) is the non-locally smeared annihilation (creation) 
operator for nucleons,
\begin{equation}
\tilde{a}_{i,j}(\boldsymbol{n})=a_{i,j}(\boldsymbol{n})+s_{\rm NL}\sum_{|\boldsymbol{n}^{\prime}-\boldsymbol{n}|=1}a_{i,j}(\boldsymbol{n}^{\prime}).
\end{equation}
and $\tilde{b}$ ($\tilde{b}^{\dagger}$) ise the non-locally smeared annihilation (creation) 
operator for hyperons,
\begin{equation}
\tilde{b}_{i}(\boldsymbol{n})= b_{i}(\boldsymbol{n}) 
                             + s_{\rm NL}\sum_{|\boldsymbol{n}^{\prime}-\boldsymbol{n}|=1}
                               b_{i}(\boldsymbol{n}^{\prime})\,,
\end{equation}
with the non-local smearing parameter $s_{\rm NL}$. We now define the purely 
locally smeared density operators with a local smearing parameter  $s_{\rm L}$ 
and a range $d$ as, 
\allowdisplaybreaks[3]
\begin{subequations}
\begin{align}
\rho^{(d)}(\boldsymbol{n}) =& \sum_{i,j=0,1} 
a^{\dagger}_{i,j}(\boldsymbol{n}) \, a^{\,}_{i,j}(\boldsymbol{n})
\nonumber\\
&+ s_{\rm L}
 \sum_{|\boldsymbol{n}-\boldsymbol{n}^{\prime}|^2 = 1}^d 
 \,
 \sum_{i,j=0,1} 
a^{\dagger}_{i,j}(\boldsymbol{n}^{\prime}) \, a^{\,}_{i,j}(\boldsymbol{n}^{\prime}) \,,\\
     \rho^{(d)}_{S}(\boldsymbol{n}) =& \sum_{i,j,i^{\prime}=0,1} 
    a^{\dagger}_{i,j}(\boldsymbol{n}) \, [\boldsymbol{\sigma}_{S}]_{i,i^{\prime}} \, a^{\,}_{i^{\prime},j}(\boldsymbol{n})
    \nonumber \\
    &+
    s_{\rm L}
     \sum_{|\boldsymbol{n}-\boldsymbol{n}^{\prime}|^2 = 1}^d 
      \sum_{i,j,i^{\prime}=0,1} 
       a^{\dagger}_{i,j}(\boldsymbol{n}^{\prime}) \, [{\boldsymbol{\sigma}}_{S}]_{ii^{\prime}} 
         \, a^{\,}_{i^{\prime},j}(\boldsymbol{n}^{\prime}) \,,
         \\
     \rho^{(d)}_{I}(\boldsymbol{n}) =& \sum_{i,j,j^{\prime}=0,1} 
    a^{\dagger}_{i,j}(\boldsymbol{n}) \, [{\boldsymbol{\tau}}_{I}]_{j,j^{\prime}} \, a^{\,}_{i,j^{\prime}}(\boldsymbol{n})
    \nonumber \\
    & +
    s_{\rm L}
     \sum_{|\boldsymbol{n}-\boldsymbol{n}^{\prime}|^2 = 1}^d 
     \,
      \sum_{i,j,j^{\prime}=0,1} 
       a^{\dagger}_{i,j}(\boldsymbol{n}^{\prime}) \, [{\boldsymbol{\tau}}_{I}]_{jj^{\prime}} 
         \, a^{\,}_{i,j^{\prime}}(\boldsymbol{n}^{\prime}) \,,
         \\
     \rho^{(d)}_{SI}(\boldsymbol{n}) =& \sum_{i,j,i^{\prime},j^{\prime}=0,1} 
    a^{\dagger}_{i,j}(\boldsymbol{n}) \,
    [{\boldsymbol{\sigma}}_{S}]_{ii^{\prime}} \, 
     [{\boldsymbol{\tau}}_{I}]_{j,j^{\prime}} \, a^{\,}_{i^{\prime},j^{\prime}}(\boldsymbol{n})
    \nonumber \\
    &\hspace{-1.5cm}+
    s_{\rm L}
     \sum_{|\boldsymbol{n}-\boldsymbol{n}^{\prime}|^2 = 1}^d 
     \,
      \sum_{i,j,i^{\prime},j^{\prime}=0,1} 
       a^{\dagger}_{i,j}(\boldsymbol{n}^{\prime}) 
       [{\boldsymbol{\sigma}}_{S}]_{ii^{\prime}} \, 
       [{\boldsymbol{\tau}}_{I}]_{j,j^{\prime}} \,  
        a^{\,}_{i,j^{\prime}}(\boldsymbol{n}^{\prime}) \,,
         \\
    \xi^{(d)}(\boldsymbol{n}) =& \sum_{i=0,1}  
         b^{\dagger}_{i}(\boldsymbol{n}) \, b^{\,}_{i}(\boldsymbol{n})   \hspace{3.5cm}
         \nonumber\\
         & + s_{\rm L}
          \sum_{|\boldsymbol{n}-\boldsymbol{n}^{\prime}|^2 = 1}^d 
          \,
          \sum_{i,j=0,1} 
         b^{\dagger}_{i}(\boldsymbol{n}^{\prime}) \, b^{\,}_{i}(\boldsymbol{n}^{\prime})\,,
         \\
     \xi^{(d)}_{S}(\boldsymbol{n}) =& \sum_{i,i^{\prime}=0,1}  
         b^{\dagger}_{i}(\boldsymbol{n}) \, [{\boldsymbol{\sigma}}_{S}]_{i,i^{\prime}} \, b^{\,}_{i^{\prime}}(\boldsymbol{n})
         \nonumber \\
         & +
         s_{\rm L}
          \sum_{|\boldsymbol{n}-\boldsymbol{n}^{\prime}|^2 = 1}^d 
          \,
           \sum_{i,i^{\prime}=0,1}  
            b^{\dagger}_{i}(\boldsymbol{n}^{\prime}) \, [{\boldsymbol{\sigma}}_{S}]_{i,i^{\prime}} 
              \, b^{\,}_{i^{\prime}}(\boldsymbol{n}^{\prime}) \,.
\end{align} 
\end{subequations}
Finally, we define
both locally and non-locally smeared density operators as, 
\begin{subequations}
\begin{align}
    \tilde{\rho}(\boldsymbol{n}) = 
    & \sum_{i,j=0,1} 
       \tilde{a}^{\dagger}_{i,j}(\boldsymbol{n}) \, \tilde{a}^{\,}_{i,j}(\boldsymbol{n})
    \nonumber  \\
    & +
    s_{\rm L}
     \sum_{|\boldsymbol{n}-\boldsymbol{n}^{\prime}|^2 = 1} 
     \,
     \sum_{i,j=0,1} 
    \tilde{a}^{\dagger}_{i,j}(\boldsymbol{n}^{\prime}) \, \tilde{a}^{\,}_{i,j}(\boldsymbol{n}^{\prime})
    \,,\\
    \tilde{\rho}_{I}(\boldsymbol{n}) = 
    & \sum_{i,j,j^{\prime}=0,1} 
    \tilde{a}^{\dagger}_{i,j}(\boldsymbol{n}) \,\left[{\boldsymbol{{\boldsymbol{\tau}}}}_{I}\right]_{j,j^{\prime}} \, \tilde{a}^{\,}_{i,j^{\prime}}(\boldsymbol{n})
    \nonumber\\
    & + s_{\rm L}
     \sum_{|\boldsymbol{n}-\boldsymbol{n}^{\prime}|^2 = 1} 
     \,
      \sum_{i,j,j^{\prime}=0,1} 
    \tilde{a}^{\dagger}_{i,j}(\boldsymbol{n}^{\prime}) \,\left[{\boldsymbol{{\boldsymbol{\tau}}}}_{I}\right]_{j,j^{\prime}} \, \tilde{a}^{\,}_{i,j^{\prime}}(\boldsymbol{n}^{\prime})
    \,,\\
    \tilde{\xi}(\boldsymbol{n}) = 
    & \sum_{i=0,1} 
       \tilde{b}^{\dagger}_{i}(\boldsymbol{n}) \, \tilde{b}^{\,}_{i}(\boldsymbol{n})
    +
    s_{\rm L}
     \sum_{|\boldsymbol{n}-\boldsymbol{n}^{\prime}|^2 = 1} 
     \,
     \sum_{i=0,1} 
    \tilde{b}^{\dagger}_{i}(\boldsymbol{n}^{\prime}) \, \tilde{b}^{\,}_{i}(\boldsymbol{n}^{\prime})
    \,.
    \end{align}
\end{subequations}
\noindent
We also note that, for our lattice MC simulations, we employ the AFQMC approach, 
which significantly suppresses sign oscillations. The general framework of 
the AFQMC method for nucleons is described in detail in Ref.~\cite{Lahde:2019npb}. 
We extend this framework to include hyperons following the approach 
of Ref.~\cite{Tong:2024egi}. It is important to note that our calculations consider only $\Lambda$ hyperons, 
as they are the most significant hyperons known to form bound states in the 
$S=-1$ sector. Although the $\Lambda$ hyperon is known to mix with the $\Sigma^0$, 
we neglect this effect in the current work but will discuss its inclusion later on.

\subsection{Hyperon-Nucleon Interactions\label{sec:YN-interaction}}

In this section we discuss the details of how to constrain the low-energy constants 
(LECs) of the $Y N$ interactions and their incorporation into 
many-body lattice calculations.

\begin{figure*}[htb]
    \centering
    \includegraphics{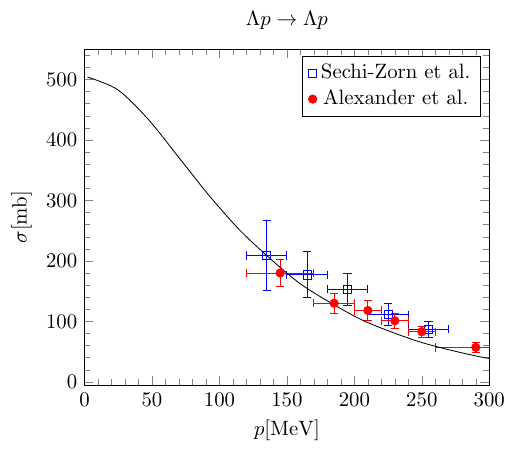}
    \includegraphics{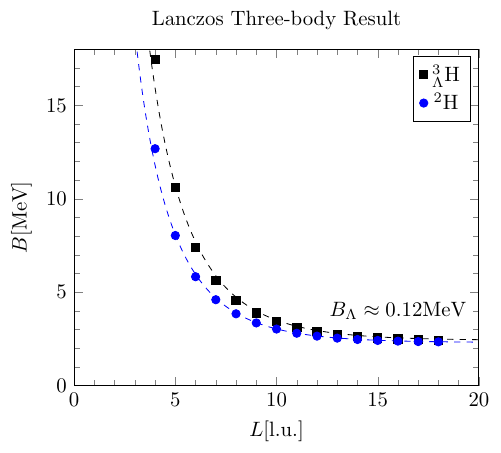}
    \caption{Left panel: Fitted $\Lambda p\to \Lambda p$ cross section \cite{Alexander:1968acu,Sechi-Zorn:1968mao}. The coupling constants are chosen to be close to the low-energy scattering parameters  of the best continuum chiral SMS N$^2$LO interaction~\cite{Haidenbauer:2023qhf}, small deviations at higher momenta are therefore expected at leading order. Right panel: Results of an exact  Lanczos  three-body calculation to confirm that the hypertriton is indeed a shallow bound state for the chosen hypernuclear interaction. The dashed lines are large $L$ extrapolations according to Eq.~\eqref{eq:Lextra}.}
    \label{fig:2body-interaction}
\end{figure*}

The LECs of $Y N$ interactions are determined 
by fitting the unpolarized cross section at infinite volume limit, while ensuring 
consistency with the scattering parameters of the best continuum interaction 
as described in Ref.~\cite{Haidenbauer:2023qhf}. Since the splitting of the two 
S-wave channels, $^1S_0$ and $^3S_1$, cannot be obtained by fitting the 
cross-section alone, we also use the bound three-body system, the hypertriton, and ensure 
it maintains a shallow bound state. It is important to note that to accurately calculate 
the shallow bound state of the hypertriton, we use an exact Lanczos code 
to eliminate stochastical errors.

Due to the large uncertainty in the empirical $\Lambda$ separation energy for the 
hypertriton, the corresponding LECs cannot be precisely determined. Nevertheless, 
the hypertriton serves as a constraint. We note that small 
variations in the $\Lambda$ separation energy of the hypertriton still allow 
for an accurate description of heavier hypernuclei~\cite{Le:2019gjp}. The results 
for the $\Lambda p \rightarrow \Lambda p $ cross section and the energies of $^2$H and $^3_{\Lambda}$H 
at finite volumes are shown in Fig.~\ref{fig:2body-interaction}.

The extracted scattering parameters from the effective range expansion are close 
to {those obtained} with semi-local momentum-space regularized (SMS) N$^2$LO interaction introduced 
in Ref.~\cite{Haidenbauer:2023qhf}. The scattering parameters obtained in 
this work are summarized in Tab.~\ref{tab:YN-scattering-parameters} and compared
with the results of Ref.~\cite{Haidenbauer:2023qhf}. For further details on the interaction, 
see Appendix~\ref{app:intdet}.
\begin{table}[htb]
    \centering
    \begin{tabular}{|l|ccc|c|}
        \hline\hline
       & \multicolumn{3}{c|}{SMS N$^2$LO~\cite{Haidenbauer:2023qhf}} & {This work} \\
       & \multicolumn{3}{c|}{$\Lambda_{\chi}$~(MeV)} &            \\
       & ~500        & ~550        & ~600               &             \\
       \hline\hline
    $a_s$ [fm] & -2.80      & -2.79      & -2.80      & -2.89            \\
    $r_s$ [fm] & ~2.82      & ~2.89      & ~2.68      & ~3.28            \\
             \hdashline
    $a_t$ [fm] & -1.56      & -1.58      & -1.56      & -1.60            \\
    $r_t$ [fm] & ~3.16      & ~3.09      & ~3.17      & ~3.94            \\
    \hline\hline                      
    \end{tabular}
    \caption{Results of scattering lengths and effective ranges for singlet and triplet S-wave 
    $\Lambda N$ scattering phase shifts and comparison with results from the SMS N$^2$LO interaction in $\chi$EFT~\cite{Haidenbauer:2023qhf}.}
    \label{tab:YN-scattering-parameters}
\end{table}

Since the hypertriton is very shallow, the $\Lambda-d$ separation is expected 
to be approximately $11$~fm \cite{Hildenbrand:2019sgp}. Therefore, typical box 
sizes used in lattice calculations with high-fidelity chiral interactions are 
not sufficient to eliminate all finite volume effects. To address this issue
we use a practical solution involving a simplified $NN$ interaction\footnote{We 
use this simplified interaction to speed up the fitting procedure. The extrapolated result 
of the N$^3$LO interaction is fully compatible with this result.} that 
enables us to use very large boxes, up to $L=24$~fm in our exact calculations. 
Additionally, we extrapolate to the infinite volume limit using the ansatz~\cite{Luscher:1985dn},
\begin{align}\label{eq:Lextra}
    E(L)=E_\infty +\frac{A}{L} \, \exp(-L/L_0) \,,
\end{align}
for both cases, the deuteron and the hypertriton. We expect this two-body formula to be valid according to Ref.~\cite{Konig:2017krd}, since the hypertriton can be subdivided into two-bound clusters, the deuteron and the $\Lambda$.  

In our lattice simulations, we follow Ref.~\cite{Tong:2024egi} and use 
the AFQMC formulation in a similar manner to include hyperons. As it is discussed in the Supplementary Sections of 
Ref.~\cite{Elhatisari:2022qfr}, in the non-perturbative part of the lattice simulations 
for nucleons a simple Hamiltonian, which consists of approximate SU(4) symmetric 
interaction, is used and it is defined as, 
\begin{align}
\begin{split}
    H^S=  & H^{N}_{\rm free} + \frac{C_{NN}}{2}\sum_{\boldsymbol{n}}:
    \left[
    \tilde{\rho}(\boldsymbol{n})
    \right]^2
    :  \\
    & + \frac{C_{NN}^{I}}{2}\sum_{I,\boldsymbol{n}}:
    \left[
    \tilde{\rho}_{I}(\boldsymbol{n})
    \right]^2
    : + 
    V_{\rm OPE} ,
    \label{eq:Hsimple}
    \end{split}
\end{align}
where $H^{N}_{\rm free}$ is the kinetic energy term for nucleons defined 
by using fast Fourier transforms to produce the exact dispersion 
relations $E_{N} =p^2/(2m_{N})$ with nucleon mass 
$m_{N}=938.92$~MeV, $C_{NN}$ is the coupling constant of the short-range SU(4) 
symmetric interaction, $C_{NN}^{I}$ is the coupling constant of the short-range isospin breaking 
interaction, $V_{\rm OPE}$ is the long-range one-pion-exchange (OPE) potential 
for nucleons, the $::$ symbol indicates normal ordering. We use local smearing parameter 
$s_{\rm L}=0.07$ and non-local smearing parameter 
$s_{\rm NL} = 0.5$ as they are set in Ref.~\cite{Elhatisari:2022qfr}. To consider 
non-perturbative contributions for the $Y N$ interactions, we define a spin-averaged 
$Y N$ interaction,
\begin{align}
    C_{Y N} = \frac{1}{4}(C_{YN}^S + 3\,C_{YN}^T)\,,
    \label{eq:DeltaV-lambdaN}
\end{align}
which enables us to derive an auxiliary field
formulation for systems consisting of nucleons and $\Lambda$ hyperons. To end this, 
we redefine the second term of Eq.~(\ref{eq:Hsimple}) as,
\begin{align}
    \frac{C_{NN}}{2}\sum_{\boldsymbol{n}}:
    \left[
    \tilde{\rho}(\boldsymbol{n})
    \right]^2
    :  
+ C_{Y N}
\sum_{\boldsymbol{n}}
   :
    \tilde{\rho}(\boldsymbol{n})
    \tilde{\xi}(\boldsymbol{n})    
    :\,.
    \label{eqn:2-baryon-int}
\end{align}
The expression given in Eq.~(\ref{eqn:2-baryon-int}) can rewritten by completing the square as
\begin{align}
    \frac{C_{NN}}{2}
  \sum_{\boldsymbol{n} }
  :
  \Ke{\tilde{\rho}\K{{\boldsymbol{n}}}+\frac{C_{Y N}}{ C_{NN}}\tilde{\xi}\K{{\boldsymbol{n}}}}^2
  :
  =
\frac{C_{NN}}{2}\sum_{\boldsymbol{n}}:
\left[
\tilde{\slashed{\rho}}(\boldsymbol{n})
\right]^2
:  
\end{align}
We note that this expression induces a $YY$ interaction, which, however, 
due to the {absence of a} second $\Lambda$ is zero by {construction}. While this modified density allows 
parallel treatment of nuclear as well as hypernuclear 
contact interactions with one auxiliary field, we observe degeneracy for the states with different 
total spin states in hypernuclei 
due to the spin-averaged $SU(4)$ symmetric interaction. 
Nevertheless, the corrections to the spin-averaged 
$SU(4)$ symmetric interaction, $\Delta V_{Y N} = V_{Y N} - \mathbbm{1}\,C_{Y N}$ are treated 
using first order 
perturbation theory, and for systems with a low-lying excited state we lift the 
degeneracy of ground and excited 
states using degenerate perturbation theory based on the nuclear ground state, see also 
appendix \ref{app:degpert}. Those separation of states within e.g. $A=4$ nucleus are 
of particular interest since they are measured very accurately and are sensitive to 
the spin of the $\Lambda$.

\subsection{Fit of the Three-Body Forces\label{sec:fit}}

In this section, we discuss the details of the three-baryon interactions $V_{YNN}$ given 
in Eq.~(\ref{eq:V-LambdaNN}) and how we regulate their singular short-distance properties. 
Recent \emph{ab-initio} calculations have shown that the locality of the short range 
interactions plays significant role on determination of the properties of atomic 
nuclei~\cite{Elhatisari:2016owd,Elhatisari:2022qfr}. 
Therefore, we construct and analyse the interactions given in Eq.~(\ref{eq:V-LambdaNN}) with 
all possible choice of smearing, both locally and non-locally. Such a procedure is not only 
important to get an accurate description for the properties of hypernuclei but also it is important to emulate the range of missing meson exchange forces. 

Hence we introduce the locally smeared forms of the interactions given 
in Eq.~(\ref{eq:V-LambdaNN}),
\begin{align}
\begin{split}
V_{1}^{(d)} = & 
3 \,
\sum_{\boldsymbol{n}}
 : \left\{[\rho^{(d)}(\boldsymbol{n})]^2 \,
    - \sum_{S} \, [\rho_{S}^{(d)}(\boldsymbol{n})]^2
    \right\}
    \xi^{(d)}(\boldsymbol{n}):
\\
  & + \sum_{\boldsymbol{n},I} 
  : \left\{ [\rho_{I}^{(d)}(\boldsymbol{n})]^2 
  - \sum_{S} \, [\rho_{SI}^{(d)}(\boldsymbol{n})]^2
  \right\}
    \xi^{(d)}(\boldsymbol{n}) : \,,
\end{split}
\\
\begin{split}
    V_{2}^{(d)} = & 
    2 \,
    \sum_{\boldsymbol{n}}:
    \rho^{(d)}(\boldsymbol{n})
    \sum_{S} \rho_{S}^{(d)}(\boldsymbol{n})
     \xi_{S}^{(d)}(\boldsymbol{n}) :
     \\
        &  -2 \, \sum_{\boldsymbol{n},S,I}:
        \rho_{I}^{(d)}(\boldsymbol{n})
        \rho_{SI}^{(d)}(\boldsymbol{n})
         \xi_{S}^{(d)}(\boldsymbol{n}): \,,
\end{split}
\\
\begin{split}
    V_{3}^{(d)} = & 
    3 \,
\sum_{\boldsymbol{n}}
: \left\{[\rho^{(d)}(\boldsymbol{n})]^2 \,
    - \sum_{I} \, [\rho_{I}^{(d)}(\boldsymbol{n})]^2 \right\}
    \xi^{(d)}(\boldsymbol{n}) :
\\
        &  + \sum_{\boldsymbol{n},S} 
        : \left\{ [\rho_{S}^{(d)}(\boldsymbol{n})]^2 
        - \sum_{I} \, [\rho_{SI}^{(d)}(\boldsymbol{n})]^2
        \right\}
          \xi^{(d)}(\boldsymbol{n}) : \,.
          \end{split}
    \label{eqn:V_c3-sL-001}  
\end{align}
Here, the superscript $d$ describes the range of the local smearing, and we consider 
different choices up to $d = 3$ which corresponds to $2.28$~fm. In addition, 
for these interactions we set $s_{\rm L} = 0.5$. Therefore, the locally smeared 
three-body interactions are labelled as {$V_{k}^{(d=0)}$, $V_{k}^{(d=1)}$, $V_{k}^{(d=2)}$ and 
$V_{k}^{(d=3)}$ with $k = 1,2,3$}. We also define the non-locally smeared forms of 
the interactions given 
in Eq.~(\ref{eq:V-LambdaNN}),
\begin{align}
\begin{split}
    V_{1}^{s_{\rm NL}} = & 
    3 \,
    \sum_{\boldsymbol{n}}
     : \left\{ [\hat{\rho}(\boldsymbol{n})]^2 \,
        - \sum_{S} \, [\hat{\rho}_{S}(\boldsymbol{n})]^2 \right\}
        \hat{\xi}(\boldsymbol{n}) :
\\
      &+ \sum_{\boldsymbol{n},I} 
      : \left\{ [\hat{\rho}_{I}(\boldsymbol{n})]^2 
      - \sum_{S} \, [\hat{\rho}_{SI}(\boldsymbol{n})]^2
      \right\}
      \hat{\xi}(\boldsymbol{n}) : \,,
    \label{eqn:V_c1-sL-001}  
    \end{split}\\
    \begin{split}
        V_{2}^{s_{\rm NL}} = & 
        2 \,
        \sum_{\boldsymbol{n}}:
        \hat{\rho}(\boldsymbol{n})
        \sum_{S} \hat{\rho}_{S}(\boldsymbol{n})
        \hat{\xi}_{S}(\boldsymbol{n}): \\
            &-2 \, \sum_{\boldsymbol{n},S,I}:
            \hat{\rho}_{I}(\boldsymbol{n})
            \hat{\rho}_{SI}(\boldsymbol{n})
            \hat{\xi}_{S}(\boldsymbol{n}): \,,      
    \end{split}
         \\
\begin{split}
        V_{3}^{s_{\rm NL}} = & 
        3 \,
    \sum_{\boldsymbol{n}}
     : \left\{[\hat{\rho}(\boldsymbol{n})]^2 \,
        - \sum_{I} \, [\hat{\rho}_{I}(\boldsymbol{n})]^2\right\}
        \hat{\xi}(\boldsymbol{n}) : \\
            &+ \sum_{\boldsymbol{n},S} 
            : \left\lbrace [\hat{\rho}_{S}(\boldsymbol{n})]^2 
            - \sum_{I} \, [\hat{\rho}_{SI}(\boldsymbol{n})]^2
            \right\rbrace
              \hat{\xi}(\boldsymbol{n}) : \,.   
\end{split}
\end{align}
Here, the superscript ${s_{\rm NL}}$ indicates the strength of the non-locality 
of the interactons. We consider three different values of the parameter 
${s_{\rm NL}}=0.1, 0.2, 0.3$, labeled {$V_{k}^{{s_{\rm NL}}=0.1}$, 
$V_{k}^{{s_{\rm NL}}=0.2}$ and $V_{k}^{{s_{\rm NL}}=0.3}$ with $k = 1,2,3$}. 
In the following, we reduce the superscript of these interaction and they are 
denoted by {$V_{k}^{0.1}$, $V_{k}^{0.2}$ and $V_{k}^{0.3}$}.

Now we consider all possible versions of the $\Lambda NN$ 
interaction, constructed from the combinations of these smeared versions 
of {$V_{1}$, $V_{2}$, and $V_{3}$}, which leads to a total set 
of $343$ combinations. By systematically analyzing each 
combination using hypernuclei from light to medium mass, we determine 
the optimal configuration for the $\Lambda NN$ 
interaction which give{s} a good description for hypernuclei. An analysis
of such kind revisiting the nuclear three-body forces
is in preparation by some of the authors.

In order to increase the predictive power of the theory we want 
to fit to a limited set of hypernuclei.
We distinguish here between two different type{s} of hypernuclei. 
Those with an $\alpha$-like nuclear core, which are insensitive to 
details of the spin-dependent force and the ones that exhibit a high 
sensitivity to spin-dependent forces by having a low-lying excited state. 
While the latter are needed to extract $C_2$, $\alpha$-like hypernuclei {are important to scale correctly towards the medium-mass region and permit to access the overall strength of $C_{1}$ and $C_{3}$}.

We do a least square fit to determine the LECs of  $\Lambda NN$  interaction. 
However, due to the lack of explicit two-pion exchange 
interactions, smeared forces emulate the long-range part of the potential 
and 
hence introduce an additional amount of freedom as mentioned before. 
Therefore, in order to estimate the quality of our description we use  
the Root Mean Square Deviation (RMSD) defined as follows
\begin{equation}\label{eq:RMSD}
    \text{RMSD}(S)=\sqrt{\frac{1}{M_S}\sum_{i\in S}\left(\frac{^{i}B^c_\Lambda-{}^{i}B^{\exp}_\Lambda}{^{i}B^{\exp}_\Lambda}\right)^2},
\end{equation}
where $^i B_\Lambda^c$ is the evaluated $\Lambda$ separation energy and 
$^i B^{\exp}_\Lambda$ is the experimental separation energy for each hypernucleus within the set.
The size of the set of hypernuclei is given by $M_S$ of the set $S$, which contains well measured hypernuclei from the light and medium mass region, starting from {$A=4$ up to $A=16$}. The corresponding $^i B^{\exp}_\Lambda$ are taken from Ref.~\cite{HypernuclearDataBase}. For 
reference the RMSD for the calculation without hypernuclear three-body 
forces is $\text{RMSD}_{\text{no $\Lambda$NN}}(S)=18.4\%$.

In order to {exemplify} how well the $YNN$ forces can 
be constrained from light hypernuclei, we consider here two scenarios. 
In scenario~$1$, we constrain the three-body forces only by the light 
$A=4$ and $A=5$ system, while in scenario~$2$ we use hypernuclei 
up to $A=16$.
Since the splitting between the ground state and the excited state in 
the hydrogen and the helium four-body system is supposedly a charge 
symmetry effect \cite{Le:2022ikc}, which we cannot resolve with the 
current setup, we take here the average.

A good starting point of choosing such hypernuclear three-body forces is decuplet 
saturation as described in Ref.~\cite{Petschauer:2020urh}. In this approach $C_3=C_1$ 
and $C_2=0$, since equal coupling strengths require that the forces are equally smeared, 
only seven combination of three-body forces remain. We start from this assumption and gradually relax the 
{constraints} to obtain sets of three-body $YNN$ forces that describe the data better and 
better. We expect that deviations from decuplet saturation are needed in the end since 
we do not yet include the explicit $\Lambda-\Sigma^0$ conversions.
The seven different combinations are listed in Tab.~\ref{tab:YNN-Decuplet} {with the best interaction pair resulting in an RMSD of $9.3\%$ while the worst has an RMSD of $14.7\%$}.

\begin{table}  \centering
    \centering
    \begin{tabular}{ c|c c}
        \hline
     {Interaction}&\multicolumn{2}{c}{RMSD$[\%]$}\\
         $V_{YNN}$ &$A=4/5$&$A\geq4$ \\
         \hline
         $V_{1}^{0.1}+V_{3}^{0.1}$&$9.3$&$9.2$\\
         $V_{1}^{0.2}+V_{3}^{0.2}$&$9.5$&$9.3$\\
         $V_{1}^{0.3}+V_{3}^{0.3}$&$9.9$&$9.7$\\
         $V_{1}^{(d = 0)}+V_{3}^{(d = 0)}$&$12.9$&$12.6$\\
         $V_{1}^{(d = 1)}+V_{3}^{(d = 1)}$&$14.7$&$14.6$\\
         $V_{1}^{(d = 2)}+V_{3}^{(d = 2)}$&$14.7$&$14.6$\\
         $V_{1}^{(d = 3)}+V_{3}^{(d = 3)}$&$14.7$&$14.6$\\
        \hline
    \end{tabular}
    \caption{Possible combination of $YNN$ forces enforcing 
    decuplet saturation $C_3=C_1$ and $C_2=0$, sorted by the RMSD fitted to the four- and 
    five-body system. The most right column gives the result fitted 
    to the complete set of hypernuclei (scenario 2).}
    \label{tab:YNN-Decuplet}
\end{table}

In the next step we try to improve the description by allowing 
the $\Lambda$ spin-dependent force, parameterized by $C_2$, 
to be non-zero. In total $21$ of the $49$ combination of forces 
improve the overall result. In order to keep the main text readable 
we only list the five combinations with the least RMSD in 
Tab.~\ref{tab:YNN-Decuplet+1}. For the full list we refer to 
Appendix~\ref{app:Fit}. Since this three-body force depends on 
the spin of the $\Lambda$, it can directly influence the splitting 
of the four-body system, and therefore we see a significant 
improvement of the RMSD. {The best interaction set has an RMSD of $5.7\%$, while the worst that still improves the interaction has an RMSD of $9.1\%$}. In addition, it confirms that our original set of non-local three-body forces is sufficient, since 
the weaker smeared force produces a better result. Thus non-local 
smearing parameters larger than ${s_{\rm NL}}=0.3$ are unlikely to 
provide a more accurate description.
\begin{table}[htb]
    \centering
    \begin{tabular}{ c|c c}
        \hline
     {Interaction}&\multicolumn{2}{c}{RMSD$[\%]$}\\
         $V_{YNN}$ &$A=4/5$&$A\geq4$ \\
         \hline
        $V^{0.1}_1+V^{(d = 0)}_2+V^{0.1}_3$&$5.7$&$5.3$\\
        $V^{0.2}_1+V^{(d = 0)}_2+V^{0.2}_3$&$5.8$&$5.4$\\
        $V^{0.3}_1+V^{(d = 0)}_2+V^{0.3}_3$&$6.3$&$5.9$\\
        $V^{0.1}_1+V^{(d = 3)}_2+V^{0.1}_3$&$7.6$&$7.4$\\
        $V^{0.2}_1+V^{(d = 3)}_2+V^{0.2}_3$&$7.7$&$7.5$\\
        \hline
    \end{tabular}
    \caption{The $5$ combinations with the least RMSD when constrained 
    via the four- and five-body systems with $C_3=C_1$. 
    A complete list can be found in Appendix~\ref{app:Fit}.}
    \label{tab:YNN-Decuplet+1}
\end{table}

In a last step we take all $343$ combinations into account and find $27$ combinations that improve the description when fitting from the $A=4$ and $A=5$ hypernuclei\footnote{Since the fit is done only for light nuclei, this new freedom might increase the overall RMSD.}. Interestingly none of those features a smeared $\Lambda$ spin-dependent force. {The best interactions show here an RMSD of $3.7\%$ for the first scenario, while the less effective are as good as the best restricted ones.} We list again the five combinations with the least RMSD in Tab.~\ref{tab:YNN-TBF}.
\begin{table}[htb]
    \centering
    \begin{tabular}{ c|c c}
        \hline
     {Interaction}&\multicolumn{2}{c}{RMSD$[\%]$}\\
         $V_{YNN}$ &$A=4/5$&$A\geq4$ \\
         \hline
        $V^{(d=2)}_1+V^{(d=0)}_2+V^{(d=1)}_3$&$3.7$&$3.6$\\
        $V^{(d=1)}_1+V^{(d=0)}_2+V^{(d=2)}_3$&$3.7$&$3.7$\\
        $V^{(d=1)}_1+V^{(d=0)}_2+V^{(d=1)}_3$&$3.8$&$3.7$\\
        $V^{(d=1)}_1+V^{(d=0)}_2+V^{(d=3)}_3$&$3.9$&$3.7$\\
        $V^{(d=1)}_1+V^{(d=0)}_2+V^{(d=0)}_3$&$4.1$&$4.1$\\
        \hline
    \end{tabular}
    \caption{The $5$ combinations with the least RMSD when 
    constrained via the four- and five-body systems.
     A complete list can be found in Appendix~\ref{app:Fit}.}
    \label{tab:YNN-TBF}
\end{table}

For this procedure we use well measured hypernuclei in the low-mass region. 
For a comprehensive overview of hypernuclear binding energies, see for 
example~\cite{HypernuclearDataBase}.
{
We also want to give a short discussion on the development of the coupling constants for the corresponding best interactions for scenario one. In case of decuplet saturation the coupling constant $C_1=C_3$ is given by $C_1=-0.00105$. Turning on the coupling constant for the second three-body force, the coupling constant $C_1$ changes to $C_1=-0.00145$, while $C_2=-0.22051$. We want to point out that the strength of these coupling constants should not be compared  directly to each other , since $C_2$ does not feature smearing. This also makes the comparison to the full result difficult, since the structure (smearing) changes completely.
We obtain $C_1=-0.01608$, $C_2=-0.27524$ and $C_3=0.01596$. We notice some cancellations between $C_1$ and $C_3$, which we naively expect to be necessary at LO, however, one should not directly compare those numbers to the ones obtained before, due to different smearings.
}
For further details on the fitting procedure see also Appendix~\ref{app:Fit}.

\subsection{Finite Volume  Effects\label{sec:finitebox}}

In addition to the previously discussed hypertriton, which exhibits 
significant sensitivity to the box size due to its extended structure 
(for details on the MC simulations extrapolation, see Appendix~\ref{app:hyp}), 
we briefly examine the dependence on the box size $L$ for another 
critical set of hypernuclei, namely the four-body system.

As the separation energies increase, we anticipate that box sizes 
typically used in nuclear calculations will suffice, since the 
$\Lambda$ binding energies begin to exceed typical nuclear binding 
energies per nucleon. This is already evident in the four-body system.

\begin{figure}[htb]
    \centering
    \includegraphics{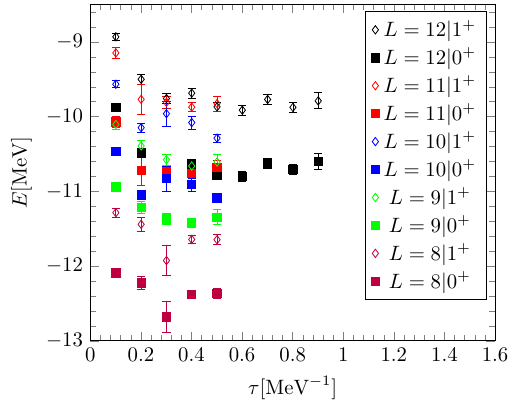}
    \caption{N$^3$LO energies for different box sizes (in units of $a$) for the excited (diamonds) as well as the ground state (squares) of ${}^4_\Lambda$H on the two-body level. The black points are used for the euclidean  time extrapolation for the final result.}
    \label{fig:boxsize}
\end{figure}

In Fig.~\ref{fig:boxsize}, we present the behavior of the $0^+$ and $1^+$ 
states in ${}^4_\Lambda$H at the two-body level. The differences 
between the smaller box sizes and those used for extrapolation with 
$L=12$ are minimal. Thus, finite box size effects are well-controlled 
for hypernuclei starting from the four-body system.

\section{Results and Discussion\label{sec:results}}

We calculate the ground state and excited state energies of hypernuclei 
up to $A = 16$. Our calculations employ high-fidelity chiral interactions at N${}^3$LO 
for nucleons, as developed in~\cite{Elhatisari:2022qfr}, leading-order S-wave hyperon-nucleon ($YN$) 
interactions constrained by the unpolarized $\Lambda p \rightarrow\Lambda p$ cross section and the hypertriton binding energy, and hyperon-nucleon-nucleon ($YNN$) interactions constrained by hypernuclear systems with $A = 4$ and $5$. For the $YNN$ interactions, we consider all possible forms of short-distance 
smearing. In our analysis, we calculate the RMSD over all calculated hypernuclear 
separation energies with $A\ge4$, which are used to assess the accuracy of the $YNN$ interactions 
in describing hypernuclei.

\begin{figure}[htb]
    \centering
    \includegraphics[width=0.485\textwidth]{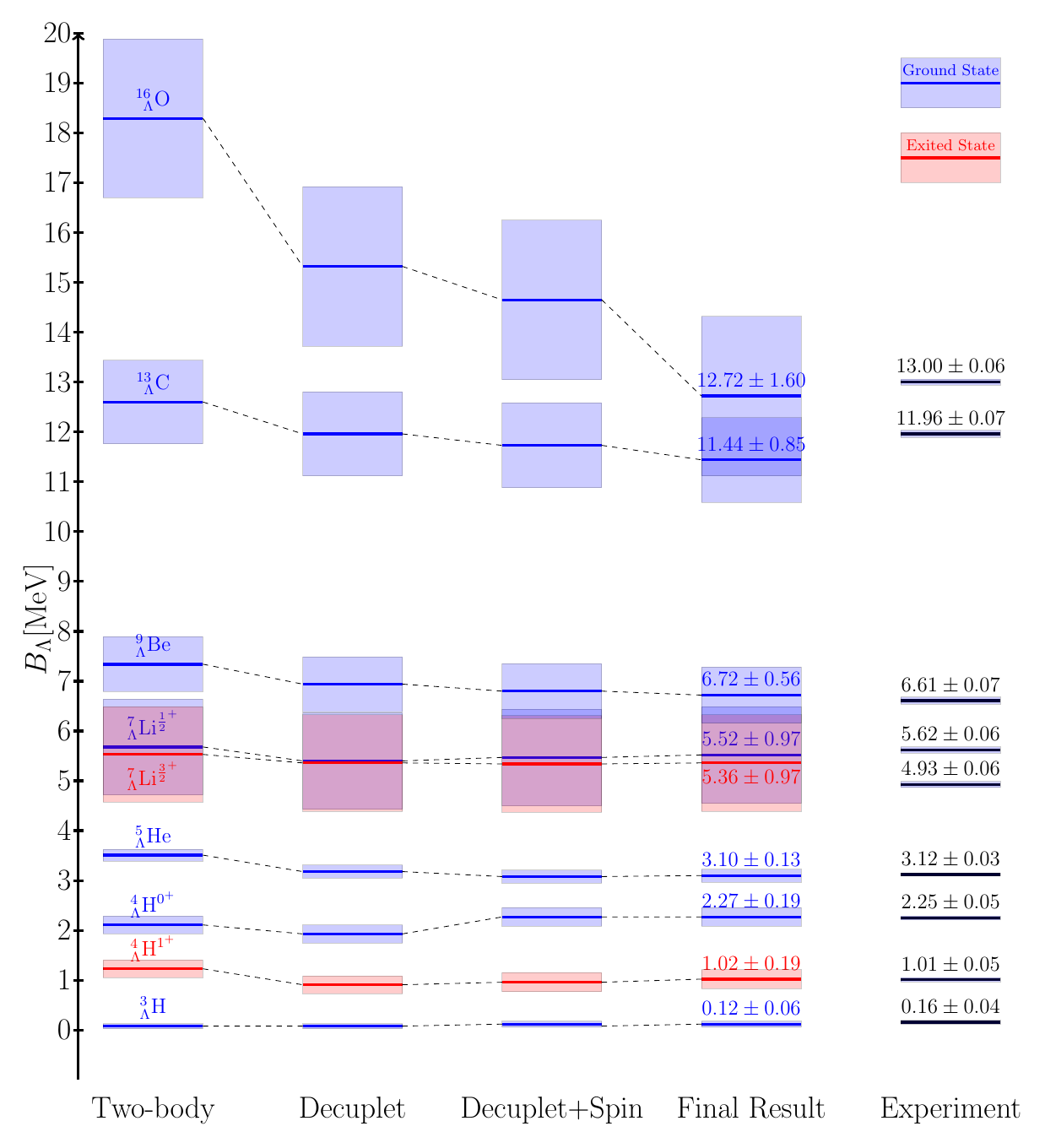}
    \caption{$\Lambda$ separation energies for different 
    $YNN$ forces in scenario~1. We choose the best combination 
    for each proposed set. The large improvement resulting from 
    the introduction of the spin-dependent three-body force is 
    clearly visible. The experimental values are taken from 
    \cite{HypernuclearDataBase}, where we averaged the four-body 
    systems. Ground states are depicted in blue, excited states in 
    red. The uncertainties are indicated by the shaded areas.}
    \label{fig:resultsS1}
\end{figure}

We present results based on only two-body $YN$ interactions, $YN$ interactions combined 
with the best set of $YNN$ interactions with decuplet approximations, and $YN$ and $YNN$ interactions  with fitted LECs
in Fig.~\ref{fig:resultsS1} and in Tab.~\ref{tab:resultsS1}. The results for $A \le 5$ shown here 
are included in the fit, while the other hypernuclei are predictions. We find that, within {stochastic uncertainties of the Monte Carlo simulations}, our Hamiltonian can accurately describe hypernuclear systems.
\begin{table*}[htb]
    \centering
    \begin{tabular}{c|c|c|c|c|c}
         Nucleus&Exp. & Two-Body(YN) & $V_{\text{Decuplet}}$ & $V_{\text{Decuplet}}\,+\,V_2\,$& Free $V_1\,+\,V_2\,+\,V_3\,$\\
         \hline
         ${}^3_\Lambda\text{H}$&$0.16\pm{0.04}$&$0.08\pm{0.05}$&$0.08\pm{0.05}$&$0.12\pm{0.06}$&$0.12\pm{0.06}$ \\
         \hdashline
         ${}^4_\Lambda\text{H}^{0^+}$&$2.25\pm{0.042}$&$2.11\pm{0.18}$&$1.93\pm{0.18}$&$2.27\pm{0.19}$&$2.258\pm{0.19}$ \\
         ${}^4_\Lambda\text{H}^{1^+}$&$1.01\pm{0.046}$&$1.23\pm{0.18}$&$0.95\pm{0.18}$&$0.97\pm{0.19}$&$1.012\pm{0.19}$ \\
         \hdashline
         ${}^5_\Lambda\text{He}$&$3.102\pm{0.03}$&$3.51\pm{0.12}$&$3.22\pm{0.12}$&$3.11\pm{0.12}$&$3.10\pm{0.13}$ \\
         \hdashline
         ${}^7_\Lambda\text{Li}^{\frac{1}{2}^+}$&$5.62\pm{0.06}$&$5.68\pm{0.96}$&$5.40\pm{0.97}$&$5.47\pm{0.97}$&$5.52\pm{0.97}$ \\
         ${}^7_\Lambda\text{Li}^{\frac{3}{2}^+}$&$4.93\pm{0.06}$&$5.53\pm{0.96}$&$5.36\pm{0.97}$&$5.34\pm{0.97}$&$5.36\pm{0.97}$ \\
         \hdashline
         ${}^9_\Lambda\text{Be}^{}$&$6.61\pm{0.07}$&$7.34\pm{0.55}$&$6.94\pm{0.56}$&$6.80\pm{0.55}$&$6.72\pm{0.55}$ \\
         ${}^{13}_{\hphantom{1}\Lambda}\text{C}^{}$&$11.96\pm{0.07}$&$12.60\pm{0.84}$&$11.96\pm{0.45}$&$11.73\pm{0.85}$&$11.44\pm{0.84}$ \\
         ${}^{16}_{\hphantom{1}\Lambda}\text{O}^{}$&$13.00\pm{0.06}$&$18.29\pm{1.59}$&$15.32\pm{1.60}$&$14.65\pm{1.60}$&$12.72\pm{1.61}$ \\
    \end{tabular}
    \caption{$\Lambda$ separation energies (in MeV) for different $YNN$ 
    forces in scenario~1. We choose the best combination 
    (least RMSD) for each proposed set. The large improvement 
    resulting from the introduction of the spin-dependent 
    three-body force is clearly visible. The experimental values 
    are taken from \cite{HypernuclearDataBase}, where we averaged 
    the four-body systems.}
    \label{tab:resultsS1}
\end{table*}

To analyze any further improvements in the results given in Tab.~\ref{tab:resultsS1}, we also perform 
fits of $YNN$ interactions and RMSD analysis by considering hypernuclei up to $A=16$. The 
results are shown in Tab.~\ref{tab:resultsS2}, and we find that the improvement in the final results is 
minimal, as the data is already well-described within uncertainties. However, the inclusion of heavier 
hypernuclei may be necessary in the future to address mid- to heavy-mass hypernuclei.
\begin{table*}[htb]
    \centering
    \begin{tabular}{c|c|c|c|c|c}
         Nucleus&Exp. & Two-Body(YN) & $V_{\text{Decuplet}}$ & $V_{\text{Decuplet}}\,+\,V_2\,$& Free $V_1\,+\,V_2\,+\,V_3\,$\\
         \hline
         ${}^3_\Lambda\text{H}$&$0.16\pm{0.04}$&$0.08\pm{0.05}$&$0.08\pm{0.05}$&$0.12\pm{0.06}$&$0.12\pm{0.06}$ \\
         \hdashline
         ${}^4_\Lambda\text{H}^{0^+}$&$2.25\pm{0.042}$&$2.11\pm{0.18}$&${1.90}\pm{0.18}$&$2.27\pm{0.19}$&$2.252\pm{0.19}$ \\
         ${}^4_\Lambda\text{H}^{1^+}$&$ 1.01\pm{0.046}$&$1.23\pm{0.18}$&$0.91\pm{0.18}$&$0.96\pm{0.19}$&$1.023\pm{0.19}$ \\
         \hdashline
         ${}^5_\Lambda\text{He}$&$3.102\pm{0.03}$&$3.51\pm{0.12}$&$3.18\pm{0.13}$&$3.08\pm{0.13}$&$3.11\pm{0.13}$ \\
         \hdashline
         ${}^7_\Lambda\text{Li}^{\frac{1}{2}^+}$&$5.62\pm{0.06}$&$5.68\pm{0.96}$&$5.35\pm{0.97}$&$5.45\pm{0.97}$&$5.51\pm{0.97}$ \\
         ${}^7_\Lambda\text{Li}^{\frac{3}{2}^+}$&$4.93\pm{0.06}$&$5.53\pm{0.96}$&$5.32\pm{0.97}$&$5.32\pm{0.97}$&$5.37\pm{0.97}$ \\
         \hdashline
         ${}^9_\Lambda\text{Be}^{}$&$6.61\pm{0.07}$&$7.34\pm{0.55}$&$6.89\pm{0.55}$&$6.76\pm{0.55}$&$6.73\pm{0.56}$ \\
         ${}^{13}_{\hphantom{1}\Lambda}\text{C}^{}$&$11.96\pm{0.07}$&$12.60\pm{0.84}$&$11.87\pm{0.84}$&$11.66\pm{0.85}$&$11.47\pm{0.85}$ \\
         ${}^{16}_{\hphantom{1}\Lambda}\text{O}^{}$&$13.00\pm{0.06}$&$18.29\pm{1.59}$&$14.86\pm{1.60}$&$14.36\pm{1.60}$&$13.00\pm{1.61}$ \\
    \end{tabular}
    \caption{$\Lambda$ separation energies (in MeV) for different $YNN$ forces in scenario~2. We choose the best combination (least RMSD) for each proposed set. The large improvement resulting from the introduction of the spin-dependent three-body force is clearly visible. The experimental values are taken from~\cite{HypernuclearDataBase}, where we averaged the four-body systems.}
    \label{tab:resultsS2}
\end{table*}

As seen from Fig.~\ref{fig:resultsS1} and Tabs.~\ref{tab:resultsS1} and ~\ref{tab:resultsS2}, 
the contributions from three-body forces exhibit the expected behavior. 
The splitting in the $A=4$ sector can be described accordingly by introducing a spin-dependent $YNN$ force. 
We also observe that the uncertainties 
in the MC simulations are dominated by operators that are only sensitive to the nuclear part of the wave function rather than the operators of the $YN$ interaction.
Additionally, the uncertainties in atomic nuclei, as described in Ref.~\cite{Elhatisari:2022qfr},
currently represent lower limits on accuracy of the separation 
energy. For the separation between the ground and excited states, the splitting 
can be determined more precisely than the results suggest due to correlated uncertainties. 
For further details, see also Appendix~\ref{app:splitting}.

For the hypertriton, we need to extrapolate to larger box sizes, and 
information on the extrapolation procedure can be found in Appendix~\ref{app:hyp}. 
Despite using a simplified nuclear interaction for the fitting procedure, 
our results are fully compatible with experimental measurements as well as our 
original fit to determine the coupling constants.
The contribution from three-body interactions is about $40$~keV.

Compared to the splitting in the $A=4$ sector, the splitting in 
the $A=7$ sector, although within uncertainties, is relatively inaccurate. 
We anticipate immediate improvement by introducing meson exchanges, 
particularly pions. This is because $\ell=2$ contributions from the nuclear core, 
which are missing in the degenerate perturbation theory construction 
based on the $1^+$ ground state, can contribute to the $\frac{3}{2}^+$ state 
(see Appendix~\ref{app:degpert}). These contributions will be then automatically included.

This leads us to possible improvements in the considered interactions here. 
We recommend including pion exchange forces in both the two-body and three-body 
sector. These forces not only allow for an automatic inclusion of higher momentum contributions 
but also make excited states available in typical multichannel calculations, as used for example
in \cite{Elhatisari:2022qfr,Shen:2021kqr,Epelbaum:2011md}, instead of relying on 
perturbation theory. Since the main contribution at the two-body level comes from 
two-pion exchange interactions rather than one-pion exchange interaction, 
we expect the sign oscillation to  be under control. 
Additionally, this approach enables the inclusion of higher orders in the chiral scheme, 
necessary for better phase shift descriptions at higher orders, including higher partial 
waves such as P-waves in the $\Lambda N$ sector, which are important for describing 
higher excited states in light hypernuclei like ${}^7_\Lambda$Li and ${}^9_\Lambda$Be.

In the final step, the explicit inclusion of the $\Sigma$ and the $\Lambda - \Sigma^0$ 
conversion could further improve the results. This inclusion is feasible in a manner 
similar to the $\Lambda$ inclusion in this work and Ref.\cite{Tong:2024egi}, or using 
a perturbative scheme as suggested in Ref.\cite{Beane:2003yx}. Such conversions are 
crucial for describing light hypernuclei within the NCSM framework \cite{Le:2023bfj,Le:2022ikc,Liebig:2015kwa,Wirth:2017lso,Wirth:2017bpw,Wirth:2016iwn,Wirth:2014apa}.

In this work we extend the successful N$^3$LO nuclear interaction towards light and medium mass hypernuclei based on a LO
\,$\Lambda N$ interaction. We present results describing accurately the ground state as well as exited states of
selected hypernuclei by including contact three-body $Y N N$ forces constrained only from the light $A=4,5$
system. This is an important step towards the {\em ab initio} description of hypernuclei in the framework of NLEFT.

\FloatBarrier

\section*{Acknowledgments}

We thank  members of the NLEFT Collaboration as well as 
Johann Haidenbauer, Andreas Nogga, Hoai Le and Hui Tong
for useful discussions.
This work was supported in part by the European
Research Council (ERC) under the European Union's Horizon 2020 research
and innovation programme (grant agreement No. 101018170),
by DFG and NSFC through funds provided to the
Sino-German CRC 110 ``Symmetries and the Emergence of Structure in QCD" (NSFC
Grant No.~11621131001, DFG Project ID 196253076 - TRR 110) and 
and the Scientific and Technological Research Council of Turkey (TUBITAK
project no. 120F341). The work of UGM was supported in part by the CAS President's International
Fellowship Initiative (PIFI) (Grant No.~2025PD0022).
The EXOTIC project is supported by the Jülich Supercomputing Centre by dedicated HPC time provided on the JURECA DC GPU partition.
The authors gratefully acknowledge the Gauss Centre for Supercomputing e.V. (www.gauss-centre.eu) 
for funding this project by providing computing time on the GCS Supercomputer JUWELS 
at J\"ulich Supercomputing Centre (JSC).

\appendix

\section{Interaction Details\label{app:intdet}}
In this work we use a spatial lattice spacing of $a=1.32$~fm and a temporal lattice spacing of $a_t=1/1000\,$MeV$^{-1}$. All interactions share a similar set of local and non-local smearing parameters of $s_{\rm L}=0.07$ and $s_{\rm NL}=0.5$, the same parameters are also used in Ref.~\cite{Elhatisari:2022qfr} for the $NN$ interaction. The presence of non-local smearing results in an 
explicit dependence on the center-of-mass momentum, thus 
breaking Galilean invariance. As it is considered for 
nucleon-nucleon interactions in Ref.~\cite{Elhatisari:2022qfr}, 
in this work we include the nucleon-hyperon Galilean Invariance Restoration (GIR) interaction. For an extensive introduction to GIR interactions on the lattice, see Ref.~\cite{Li:2019ldq}.
The LECs of the $YN$ interactions are given in Tab.~\ref{tab:intpar}.
\begin{table}[t!]
    \centering
    \begin{tabular}{|l|c|c|}
    \hline
            & $^1S_0(YN)$&$^3S_1(YN)$\\
            \hline
            $C$&$-3.16\times10^{-3}$&$-2.39\times10^{-3}$\\
            $C_\text{GIR}^0$&$-4.36\times10^{-4}$&$-3.28\times10^{-4}$\\
            $C_\text{GIR}^1$&$\hphantom{-}8.95\times10^{-5}$&$\hphantom{-}6.73\times10^{-5}$\\
            $C_\text{GIR}^2$&$-8.40\times10^{-6}$&$-6.31\times10^{-6}$\\
            \hline
    \end{tabular}
    \caption{Coupling constants of $YN$ interactions as well as the corresponding GIR interactions (in lattice units).}
    \label{tab:intpar}
\end{table}
In the evolution in our MC simulation we than use the spin averaged force $C_{\text{YN}}=-2.58\times10^{-3}$.

\section{Construction Based on Degenerate Perturbation Theory\label{app:degpert}}

As discussed in Sec.~\ref{sec:YN-interaction}, we construct a unified 
{\it ab initio} theory to study hypernuclei building upon the {\it ab initio} 
theory for nuclei developed in Ref.~\cite{Elhatisari:2022qfr}. By
incorporating hyperons into the theory for nucleons, we employ a recently 
formulated AFQMC method~\cite{Tong:2024egi}, which considers only spin and isospin symmetric 
$YN$ interaction in the non-perturbative calculations. Such a choice enables 
efficient calculations for the systems consisting of nucleons and hyperons, but 
at the cost of observing degeneracy in the low-lying energy levels for some 
hypernuclei, such as $^{4}_{\Lambda}$H and  $^{7}_{\Lambda}$Li. To lift such type of degeneracy 
we treat the spin breaking $YN$ and $YNN$ 
interactions using degenerate perturbation theory (DPT), an 
essential technique for accurately describing the behavior of systems with 
overlapping energy levels. In DPT, the process involves constructing the perturbation 
Hamiltonian 
within the degenerate subspace and then diagonalizing it to find the new energy 
eigenvalues and corresponding eigenstates.

For instance, we consider the $^{4}_{\Lambda}$H hypernucleus 
where we start with the nuclear ground state 
(e.g., the triton, which is a $J=1/2$ state due to the Pauli 
exclusion principle between nucleons) and add an additional $\Lambda$. 
To address the two lowest-lying energy states, we perform 
a multi-state calculation with quantum numbers
$J=0, J_z=0$ and $J=1, J_z=0$ using  
the following initial wave functions for the $^{4}_{\Lambda}$H 
system,
\begin{subequations}
    \begin{align}
        \ket{\psi_1}&=a^{\dagger}_{\uparrow,p}(0)a^{\dagger}_{\uparrow,n}(0)a^{\dagger}_{\downarrow,n}(0)b^{\dagger}_{\downarrow}(0)\ket{0}~,\\
        \ket{\psi_2}&=a^{\dagger}_{\downarrow,p}(0)a^{\dagger}_{\downarrow,n}(0)a^{\dagger}_{\uparrow,n}(0)b^{\dagger}_{\uparrow}(0)\ket{0}~,
    \end{align}
    \end{subequations}
where the creation operators the ones defined in the main text.

We then compute the corresponding $2\times2$ perturbation Hamiltonian 
for any operator $\mathcal{O}$ involving hyperons,
\begin{align}
H_{\mathcal{O}}^{\tau}
=
\begin{pmatrix}
\matrixel{\psi_1^{\tau}}{\mathcal{O}}{\psi_1^{\tau}}&\matrixel{\psi_1^{\tau}}{\mathcal{O}}{\psi_2^{\tau}}\\
\matrixel{\psi_2^{\tau}}{\mathcal{O}}{\psi_1^{\tau}}&\matrixel{\psi_2^{\tau}}{\mathcal{O}}{\psi_2^{\tau}}
\end{pmatrix}\,,
\label{eqn:pert-Hamil-4b}
\end{align}
where $\psi_{1,2}^{\tau}$ is the projected state at a given Euclidean time $\tau$. Due to the absence of any 
meson-exchange interaction between nucleons and 
hyperons, the relations between the entries of the perturbation 
Hamiltonian given in Eq.~(\ref{eqn:pert-Hamil-4b}), up to a stochastic error, 
can be written as
$\matrixel{\psi_1^{\tau}}{\mathcal{O}}{\psi_1^{\tau}} \approx \matrixel{\psi_2^{\tau}}{\mathcal{O}}{\psi_2^{\tau}}$ and 
$\matrixel{\psi_1^{\tau}}{\mathcal{O}}{\psi_2^{\tau}} \approx \matrixel{\psi_2^{\tau}}{\mathcal{O}}{\psi_1^{\tau}}$. As a result, the diagonalization of Eq.~(\ref{eqn:pert-Hamil-4b}) yields 
the eigenenergies for $^{4}_{\Lambda}$H$^{1^+}$ and  $^{4}_{\Lambda}$H$^{0^+}$ with 
the corresponding eigenstates 
\begin{subequations}
\begin{align}
\ket{\psi_{{}^{4}_{\Lambda}{\rm H}^{0^+}}}&=\frac{\ket{\psi_1^{\tau}}-\ket{\psi_2^{\tau}}}{\sqrt{2}}~,\\
\ket{\psi_{{}^{4}_{\Lambda}{\rm H}^{1^+}}}&=\frac{\ket{\psi_1^{\tau}}+\ket{\psi_2^{\tau}}}{\sqrt{2}}~.
\end{align}
\end{subequations}

We apply a similar procedure for the $A=7$-body system, 
although the nuclear core in this case is an $\ell=1$ state. 
Since this method is based solely on the nuclear ground state, 
contributions from excited nuclear states (which can particularly 
affect excited states in the $A=7$ system) are missing. This limits 
our predictive power for such systems. In future studies, a full 
inclusion of meson-exchange will remove the dependence on this type 
of perturbative calculation for excited states.

\section{Fitting Procedure}\label{app:Fit}

In this section, we present a comprehensive analysis of the $YNN$ interactions 
by considering all possible combinations of the interactions out of  
$\lbrace V_{1}^{(d)}, V_{1}^{s_{\rm NL}}\rbrace$, 
$\lbrace V_{2}^{(d^{\prime})}, V_{2}^{s_{\rm NL}^{\prime}}\rbrace$ 
and $\lbrace V_{3}^{(d^{\prime\prime})}, V_{3}^{s_{\rm NL}^{\prime\prime}}\rbrace$. 
Such a detailed analysis is feasible because the $YNN$ interactions are treated perturbatively 
 in our calculations.

In our simulations, we compute the derivatives of the energy of the Hamiltonian, 
as given in  Eq.~(\ref{eq:H-001}), with respect to the LECs
$C_{1}$, $C_{2}$, and $C_{3}$, that is $\partial\langle{E}\rangle/\partial C_{k}$. 
This approach allows 
us to determine the contributions from the $YNN$ interactions as perturbative corrections,
\begin{align}
\langle{E_{YNN}\rangle} = C_{1} \frac{\partial\langle{E}\rangle}{\partial C_{1}}
+C_{2} \frac{\partial\langle{E}\rangle}{\partial C_{2}}
+C_{3} \frac{\partial\langle{E}\rangle}{\partial C_{3}}.
\end{align}
To determine the LECs $C_{1}$, $C_{2}$, and $C_{3}$, 
we perform standard least-squares fitting constrained by the separation energies. 
The results for the decuplet saturation scenario with one additional parameter 
$C_{2}$ are listed in Tab.~\ref{tab:YNN-Decuplet+1_comp_app}. 
The complete set of $YNN$ interactions without any constraints 
is provided in Tab.~\ref{tab:YNN-TBF_app}. 
For clarity, the five combinations that result in the smallest RMSD are 
highlighted in the main text  in 
Tabs.~\ref{tab:YNN-Decuplet+1} and \ref{tab:YNN-TBF}.
\begin{table}[htb]
    \centering
    \begin{tabular}{ c|c c}
        \hline
     {Interaction}&\multicolumn{2}{c}{RMSD$[\%]$}\\
         $V_{YNN}$ &$A=4/5$&$A\geq4$ \\
         \hline
        $V^{0.1}_1+V^{(d=0)}_2+V^{0.1}_3$&$5.7$&$5.3$\\
        $V^{0.1}_1+V^{(d=1)}_2+V^{0.1}_3$&$7.8$&$7.6$\\
        $V^{0.1}_1+V^{(d=2)}_2+V^{0.1}_3$&$7.7$&$7.5$\\
        $V^{0.1}_1+V^{(d=3)}_2+V^{0.1}_3$&$7.6$&$7.4$\\
        $V^{0.1}_1+V^{0.1}_2+V^{0.1}_3$&$8.7$&$8.1$\\
        $V^{0.1}_1+V^{0.2}_2+V^{0.1}_3$&$8.7$&$8.1$\\
        $V^{0.1}_1+V^{0.3}_2+V^{0.1}_3$&$8.7$&$8.1$\\

        $V^{0.2}_1+V^{(d=0)}_2+V^{0.2}_3$&$5.8$&$5.4$\\
        $V^{0.2}_1+V^{(d=1)}_2+V^{0.2}_3$&$7.9$&$7.6$\\
        $V^{0.2}_1+V^{(d=2)}_2+V^{0.2}_3$&$7.8$&$7.6$\\
        $V^{0.2}_1+V^{(d=3)}_2+V^{0.2}_3$&$7.7$&$7.5$\\
        $V^{0.2}_1+V^{0.1}_2+V^{0.2}_3$&$8.8$&$8.2$\\
        $V^{0.2}_1+V^{0.2}_2+V^{0.2}_3$&$8.8$&$8.2$\\
        $V^{0.2}_1+V^{0.3}_2+V^{0.2}_3$&$8.8$&$8.2$\\

        $V^{0.3}_1+V^{(d=0)}_2+V^{0.3}_3$&$6.3$&$5.9$\\
        $V^{0.3}_1+V^{(d=1)}_2+V^{0.3}_3$&$8.3$&$8.0$\\
        $V^{0.3}_1+V^{(d=2)}_2+V^{0.3}_3$&$8.2$&$7.9$\\
        $V^{0.3}_1+V^{(d=3)}_2+V^{0.3}_3$&$8.1$&$7.9$\\
        $V^{0.3}_1+V^{0.1}_2+V^{0.3}_3$&$9.1$&$8.5$\\
        $V^{0.3}_1+V^{0.2}_2+V^{0.3}_3$&$9.1$&$8.5$\\
        $V^{0.3}_1+V^{0.3}_2+V^{0.3}_3$&$9.1$&$8.5$\\
        \hline
    \end{tabular}
    \caption{Combination that improve the least RMSD when constrained via the four- and five-body systems with $C_3=C_1$ compared to the best fit considering decuplet saturation.
    }
    \label{tab:YNN-Decuplet+1_comp_app}
\end{table}
\begin{table}[htb]
    \centering
    \begin{tabular}{ c|c c}
        \hline
     {Interaction}&\multicolumn{2}{c}{RMSD$[\%]$}\\
         $V_{YNN}$ &$A=4/5$&$A\geq4$ \\
         \hline
        $V^{(d=0)}_1+V^{(d=0)}_2+V^{(d=1)}_3$&$4.2$&$4.0$\\
        $V^{(d=0)}_1+V^{(d=0)}_2+V^{(d=2)}_3$&$4.1$&$3.9$\\
        $V^{(d=0)}_1+V^{(d=0)}_2+V^{(d=3)}_3$&$4.6$&$4.0$\\

        $V^{(d=1)}_1+V^{(d=0)}_2+V^{(d=0)}_3$&$4.1$&$4.1$\\
        $V^{(d=1)}_1+V^{(d=0)}_2+V^{(d=1)}_3$&$3.8$&$3.7$\\
        $V^{(d=1)}_1+V^{(d=0)}_2+V^{(d=2)}_3$&$3.7$&$3.7$\\
        $V^{(d=1)}_1+V^{(d=0)}_2+V^{(d=3)}_3$&$3.9$&$3.7$\\

        $V^{(d=2)}_1+V^{(d=0)}_2+V^{(d=1)}_3$&$3.7$&$3.6$\\
        $V^{(d=2)}_1+V^{(d=0)}_2+V^{(d=2)}_3$&$4.1$&$3.6$\\
        $V^{(d=2)}_1+V^{(d=0)}_2+V^{(d=3)}_3$&$4.5$&$3.7$\\

        $V^{(d=3)}_1+V^{(d=0)}_2+V^{(d=1)}_3$&$4.2$&$3.7$\\
        $V^{(d=3)}_1+V^{(d=0)}_2+V^{(d=2)}_3$&$4.7$&$3.8$\\
        $V^{(d=3)}_1+V^{(d=0)}_2+V^{(d=3)}_3$&$5.2$&$3.8$\\
        $V^{(d=3)}_1+V^{(d=0)}_2+V^{0.1}_3$&$5.5$&$4.9$\\
        $V^{(d=3)}_1+V^{(d=0)}_2+V^{0.2}_3$&$5.5$&$4.9$\\

        $V^{0.1}_1+V^{(d=0)}_2+V^{(d=0)}_3$&$5.2$&$4.4$\\
        $V^{0.1}_1+V^{(d=0)}_2+V^{(d=1)}_3$&$4.7$&$4.0$\\
        $V^{0.1}_1+V^{(d=0)}_2+V^{(d=2)}_3$&$4.3$&$3.8$\\
        $V^{0.1}_1+V^{(d=0)}_2+V^{(d=3)}_3$&$4.2$&$3.7$\\

        $V^{0.2}_1+V^{(d=0)}_2+V^{(d=0)}_3$&$5.5$&$4.5$\\
        $V^{0.2}_1+V^{(d=0)}_2+V^{(d=1)}_3$&$4.9$&$4.1$\\
        $V^{0.2}_1+V^{(d=0)}_2+V^{(d=2)}_3$&$4.5$&$3.8$\\
        $V^{0.2}_1+V^{(d=0)}_2+V^{(d=3)}_3$&$4.3$&$3.8$\\

        $V^{0.3}_1+V^{(d=0)}_2+V^{(d=0)}_3$&$5.7$&$4.6$\\
        $V^{0.3}_1+V^{(d=0)}_2+V^{(d=1)}_3$&$5.0$&$4.1$\\
        $V^{0.3}_1+V^{(d=0)}_2+V^{(d=2)}_3$&$4.7$&$3.9$\\
        $V^{0.3}_1+V^{(d=0)}_2+V^{(d=3)}_3$&$4.5$&$3.8$\\

        \hline
    \end{tabular}
    \caption{Combinations of YNN-forces constrained via the four- and five-body systems that reduce the RMSD when compared with the best combination given in Tab~\ref{tab:YNN-Decuplet+1}, by lifting all constraints.} 
    \label{tab:YNN-TBF_app}
\end{table}

\section{\texorpdfstring{${}^3_\Lambda$H}{Hypertriton} Infinite Volume Extrapolation\label{app:hyp}}
The infinite volume extrapolation of the ${}^3_\Lambda$H energies 
from a simplified interaction is presented in the main text, 
and in this section we discuss the infinite volume extrapolation of 
the ${}^3_\Lambda$H energies using a high-fidelity interaction. 

We compute the ground state energies of the ${}^3_\Lambda$H 
system using the Hamiltonian $H_{\rm N^3LO} + H^{Y}_{\rm free} + V_{Y N}$ through 
lattice Monte Carlo simulations in box sizes from $L=8$ to $L=15$. 
In addition, we compute the deuteron binding energies using 
the Hamiltonian $H_{\rm N^3LO}$ with exact diagonalization methods.

\begin{figure}[htb]
    \centering
    \includegraphics{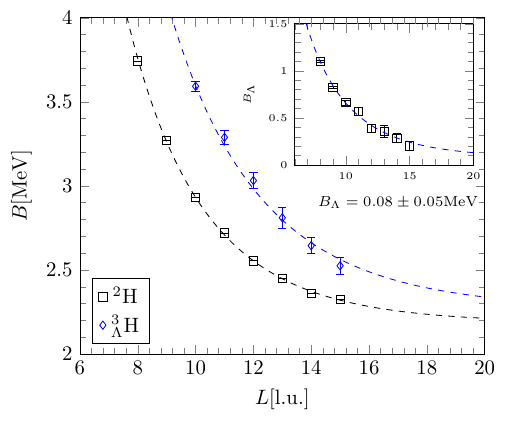}
    \caption{Infinite volume extrapolation of the ground state energies of the deuteron and the hypertriton 
    from high-fidelity interactions. The inset shows the extrapolation 
    of the separation energy.}
    \label{fig:BoxHyp}
\end{figure}

We perform infinite volume extrapolations for both the individual 
ground state energies and the $\Lambda$ separation energy simultaneously, 
employing the ansatz given in Eq.~\eqref{eq:Lextra}. 
The results are illustrated in Fig.~\ref{fig:BoxHyp}. 
The main figure shows the individual energy extrapolations, 
while the inset highlights the separation energies.

As shown in  Tabs.~\ref{tab:resultsS1} and \ref{tab:resultsS2}, 
the impact of the decuplet-saturated $YNN$ interactions 
is below $1$~keV, while the total contribution of the $YNN$ interactions 
is approximately $42$~keV. To improve the accuracy of these calculations, 
we plan to employ exact diagonalization methods for three-body and four-body hypernuclei in  our future work.

\section{Splitting between \texorpdfstring{${}^4_\Lambda\text{H}^{0^+}$}{} and \texorpdfstring{${}^4_\Lambda\text{H}^{1^+}$}{}}\label{app:splitting}

The splitting between ${}^4_\Lambda\text{H}^{0^+}$ and ${}^4_\Lambda\text{H}^{1^+}$, 
as well as the separation between ${}^4_\Lambda\text{He}^{0^+}$ and ${}^4_\Lambda\text{He}^{1^+}$, 
was initially measured with high precision using $\gamma$ spectroscopy~\cite{CERN-Lyon-Warsaw:1979ifx}. 
These measurements indicated only a small charge symmetry breaking effect. 
More recent and precise measurements of the ${}^4_\Lambda\text{He}$ system~\cite{J-PARCE13:2015uwb} 
have refined our understanding of this charge symmetry effect. Given that our 
calculations do not resolve this difference, we use the averaged value for 
this splitting, $1.247 \pm 0.010$~MeV.

Using degenerate perturbation theory, we access the splitting between 
the ground and excited states in the four-body system. This method allows 
us to extract the splitting between these two states more precisely than 
what is reported in the main results section, as the uncertainties 
from the nuclear part drop out when focusing solely on the spin-breaking $Y N$ interaction.

\begin{figure}[htb]
    \centering
    \includegraphics{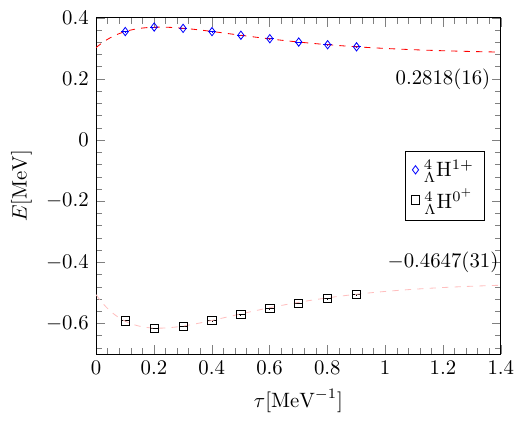}
    \caption{Contribution of the spin-dependent part of the YN interaction to the two four-body energy levels in DPT on a two-body level if extracted independently from the nuclear part. The dashed lines indicate the Euclidean time extrapolations, the difference of those contributions results in the splitting between the ground and excited state.}
    \label{fig:splitting}
\end{figure}

Fig.~\ref{fig:splitting} shows the contribution of the spin-dependent part of the YN interaction to these two states when 
the Euclidean time extrapolation of the spin-dependent two-body $Y N$ force 
is performed independently 
from the nuclear part. Since this is the only spin-dependent part on the two-body level, the difference in the contributions result in the splitting of $B_\Lambda({}^4_\Lambda\text{H}^{0^+})$-$B_\Lambda({}^4_\Lambda\text{H}^{1^+}) = 0.7465 \pm 0.0035$~MeV of the $0^+$ and $1^+$ state. 
The relatively large difference of approximately $~130$~keV, compared to the results presented in Tabs.~\ref{tab:resultsS1} 
and \ref{tab:resultsS2}, is due to the dominant influence of the nuclear part during the Euclidean time extrapolation of the main result.

\end{document}